# Relaxation, decoherence and steady-state population inversion in qubits doubly dressed by microwave and radiofrequency fields

A P Saiko[1], R Fedaruk[2], S A Markevich[1]

[1] Scientific-Practical Materials Research Centre NAS of Belarus, Minsk, Belarus

[2] Institute of Physics, University of Szczecin, 70-451, Szczecin, Poland

E-mail: saiko@ifttp.bas-net.by; fedaruk@wmf.univ.szczecin.pl

The coherent dynamics of relaxing spin qubits driven by a classical bichromatic field comprising a strong resonant component and a weaker component with a frequency close to the strong-field Rabi frequency is studied. The double dressing by the bichromatic field modifies dephasing and dissipation processes. We demonstrate that detuning of the weaker-field frequency from the strong-field Rabi frequency prolongs the decay of Rabi oscillations between some doubly dressed states. The sensitivity of Rabi oscillations to the modified detuning-dependent relaxation is illustrated for nitrogen-vacancy qubits in diamond. We discuss a steady-state population inversion of the doubly dressed-state levels.

## I. Introduction

The resonant interaction between a monochromatic electromagnetic field and a two-level quantum system (qubit) can be described in terms of dressed states [1]. The dressing of qubit by the classical electromagnetic field splits of each its level into two giving rise to two new qubits. The states of these new qubits describe composites of electromagnetic field and mater. The splitting of each level is determined by the field-qubit coupling and is characterized by the generalized Rabi frequency $\omega_R = \sqrt{\omega_{R0}^2 + \Delta^2}$, where $\Delta = \omega_0 - \omega_d$ is the detuning of the driving frequency $\omega_d$ from the resonant frequency $\omega_0$ of the qubit, and $\omega_{R0}$ is the field-qubit coupling or the Rabi frequency at the resonant ($\Delta = 0$) excitation [2]. Direct observation of the Rabi oscillations in time-resolved experiments or the observation of the Mollow triplet at frequencies $\omega_d$ and $\omega_d \pm \omega_R$ [3] in spectral pump-probe experiments constitutes a signature of the existence of the dressed states. In particular, the dressed states allow increasing the coherence time of qubits. Recently it was shown that a decrease in the damping rate of Rabi oscillations with an increase in the detuning can be observed [4,5], e.g., for artificial atoms such as semiconductor quantum dots, for which pure dephasing processes are almost absent [6,7].

Driving of the dressed qubits by the second field results in their double dressing. Such dressing of qubits can be realized by two techniques using the bichromatic field with either close or strongly different frequencies. In both of these techniques, the strong resonant field excites the qubit resulting in its primary dressing. In one case, the frequency of the second, weaker off-resonance field is detuned from the qubit resonance (the frequency of the first field) to the value close to the strong-field Rabi frequency $\omega_R$ [8,9]. In the another case, the frequency of the second, low-frequency field is close to $\omega_R$ [10-12]. Such bichromatic driving doubly dresses qubits and splits each of their initial energy states into four. The multiphoton transitions between the doubly dressed states in the coupled "qubit-bichromatic field" system have been observed in spectral and time-resolved experiments. Doubly dressed states of qubits have been studied in a wide range of physical objects, including, among others, electron and nuclear spins [10,11,13], quantum dots [14,15], superconducting qubits [16], and cold atoms [17]. The splitting of each peak of the initial Mollow triplet into three new ones has been demonstrated at the continuous-wave optical bichromatic excitation of two-level atoms [9]. The additional Rabi oscillations caused by transitions between doubly dressed states have been observed at the pulse bichromatic excitation of spin qubits in electron paramagnetic resonance (EPR) [10-12] and nuclear magnetic resonance (NMR) [13], atoms in the optical range [8], and a single nitrogen-vacancy spin qubit [18]. The bichromatic control of Rabi oscillations between doubly dressed states and prolongation of their coherence can find applications in quantum information processing. The double dressing of spin qubit is also suitable for emulation of a hybrid spin-mechanical system that has been i investigated recently to reveal the quantum behavior of matter at the macroscopic scale [18]. Moreover, in a single-spin magnetometry, Rabi oscillations between doubly dressed states open the possibility for the direct and sensitive detection of weak radio-frequency (RF) magnetic fields [19]. On the other hand, the interaction of the qubit with photons of two frequencies has a potential usage in a quantum amplifier or an attenuator [20]. The probe field is attenuated at its frequency $\omega_p \approx \omega_d + \omega_R$, whereas it is amplified at $\omega_p \approx \omega_d - \omega_R$. Amplification (lasing) is due to population inversion in the dressed-state basis and can take place without population inversion in the bare-state basis. Such amplification was demonstrated for the RF fields and nuclear spins in an optically detected nuclear magnetic resonance [21] as well as for optical fields and atoms [22]. In addition to an ensemble of qubits, amplification due to dressed-state inversion can be realized by using a single qubit [23-25]. Recently, the dressed-state amplification and attenuation of a microwave (MW) probe signal by a single superconducting qubit has been observed [20,26]. In these studies of lasing, the double dressing by the probe field was used to detect population inversion created by the driving field. However, when the probe field is stronger than the relaxation processes, the double dressing results in the observed splitting of singly dressed states and can modify the relaxation between these states. To our knowledge the features of such "doubly-dressed relaxation" have not been discussed in literature.

In the present paper, we theoretically investigate the peculiarities of dephasing and dissipation processes in the qubit driven by transverse MW and longitudinal RF fields. In Sec. II, the Liouville equation with the relaxation superoperator describing dephasing and dissipation processes in the qubit is presented. Transformations to doubly dressed states of the qubit in the strong-field limit allow us to obtain the analytical expression for the density matrix. We find that the relaxation rates of the doubly dressed states are dependent on the bichromatic field parameters. In Sec. III, we employ the newly obtained results in describtion of the decay time of Rabi oscillations. We demonstrate the main features of the "doubly-dressed relaxation" for a realistic system (nitrogen-vacancy qubits in diamond). A steady-state population inversion of the doubly dressed-state levels is the subject of Sec. IV.

**II. Doubly-dressed spin states and their relaxation and dephasing**

We consider a spin qubit with the ground $|1\rangle$ and excited $|2\rangle$ states in three fields: a MW one oriented along the $x$ axis of the laboratory frame together with RF and static magnetic fields, both directed along the $z$ axis. The Hamiltonian of the qubit can be written as

$$H = H_0 + H_\perp(t) + H_\parallel(t), \tag{1}$$

where $H_0 = \omega_0 s^z$ is the Hamiltonian of the Zeeman energy of the qubit in the static magnetic field, whereas $H_\perp(t) = \omega_1 (s^+ + s^-)\cos\omega_{mw}t$ and $H_\parallel(t) = 2\omega_2 s^z \cos(\omega_{rf}t + \psi)$ are the Hamiltonians of the qubit interaction with linearly polarized MW and RF fields, respectively. Here $\omega_{mw}$ and $\omega_{rf}$ are the frequencies of the MW and RF fields, $\psi$ is the phase of the RF field, the MW phase being set to zero. Moreover, $\omega_1$ and $\omega_2$ denote the respective interaction constants, whereas $s^{\pm,z}$ are components of the spin operator, describing the state of the qubit and satisfying the commutation relations: $[s^+, s^-] = 2s^z$, $[s^z, s^\pm] = \pm s^\pm$. We assume that $\omega_1/\omega_{mw} \ll 1$ and use the rotating-wave approximation (RWA) for the interaction between the qubit and the MW field. The case of strong driving where the RWA breaks down was reviewed e. g. in [27]. We also assume that $\omega_{mw} \gg \omega_{rf}$ and $\omega_1 \gg \omega_2$.

The dynamics of the qubit is described by the Liouville equation for the density matrix $\rho$

$$i\hbar \frac{\partial \rho}{\partial t} = [H, \rho] + i\Lambda\rho \tag{2}$$

(in the following we take $\hbar=1$). The superoperator $\Lambda$ describing decay processes is defined as

$$\Lambda\rho = \frac{\gamma_{21}}{2}D[s^-]\rho + \frac{\gamma_{12}}{2}D[s^+]\rho + \frac{\eta}{2}D[s^z]\rho, \qquad (3)$$

where $D[O]\rho = 2O\rho O^+ - O^+O\rho - \rho O^+O$, $\gamma_{21}$ and $\gamma_{12}$ are the rates of the transitions from the excited state 2 of the qubit to its ground state 1 and vice versa, and $\eta$ is the dephasing rate. The characteristic time of the coherent dynamics of qubits is considered to be much larger than the correlation times of excitations (photons, phonons) in a thermostat. Therefore, we restrict our consideration to the markovian limit by writing the master equation for the density matrix in the Lindblad form.

After two canonical transformations $\rho_2 = u_2^+u_1^+\rho u_1 u_2$, where $u_1 = \exp(-i\omega_{mw}ts^z)$ and $u_2 = \exp(-i\theta s^y/2)$, equation (3) is transformed into $i\partial\rho_2/\partial t = [H_2,\rho_2] + i\Lambda_2\rho_2$. The first canonical transformation $u_1$ moves to the frame rotating with the mw frequency $\omega_{mw}$ and eliminates the time in the part of the Hamiltonian $H_0 + H_\perp(t)$, which describes the interaction between the qubit and the mw field. The second transformation $u_2$ is used to diagonalize the Hamiltonian obtained after the first transformation $\Delta s^z + (\omega_1/2)(s^+ + s^-)$, where $\Delta = \omega_0 - \omega_{mw}$ (see also [28] – [30]). Simultaneously, $u_2$ transforms the relaxation superoperator, which retains in the RWA the Lindblad form ($\Lambda_2$), but the relaxation rates are modified. So, the second transformation realizes the transition to the singly dressed states and dressed relaxation rates. We obtain

$$H_2 = \Omega s^z - \omega_2 \sin\theta \cos(\omega_{rf} t + \psi)(s^+ + s^-) + 2\omega_2 \cos\theta \cos(\omega_{rf} t + \psi)s^z, \qquad (4)$$

$$\Lambda_2\rho_2 = \frac{\Gamma_\downarrow}{2}D[s^-]\rho_2 + \frac{\Gamma_\uparrow}{2}D[s^+]\rho_2 + \frac{\Gamma_\varphi}{2}D[s^z]\rho_2, \qquad (5)$$

$$\Gamma_\downarrow = \frac{1}{4}(\gamma_{21} + \gamma_{12})(1 + \cos^2\theta) + \frac{1}{2}(\gamma_{21} - \gamma_{12})\cos\theta + \frac{1}{4}\eta\sin^2\theta,$$

$$\Gamma_\uparrow = \frac{1}{4}(\gamma_{21} + \gamma_{12})(1 + \cos^2\theta) - \frac{1}{2}(\gamma_{21} - \gamma_{12})\cos\theta + \frac{1}{4}\eta\sin^2\theta,$$

and

$$\Gamma_\varphi = \eta\cos^2\theta + (\gamma_{21} + \gamma_{12})\sin^2\theta.$$

Moreover, $\Gamma_{\downarrow,\uparrow,\varphi}$ are the singly-dressed relaxation rates of the qubit states, $\Omega = (\omega_1^2 + \Delta^2)^{1/2}$ is the generalised Rabi frequency in the MW field, $\cos\theta = \Delta/\Omega$, and $\sin\theta = \omega_1/\Omega$. These dressed relaxation rates arise due to taking into account the strong interaction between the qubit and the MW field. Since $\gamma_{21}, \gamma_{12}, \eta \ll \omega_{rf}, \Omega$, the RWA is used to obtain the operator (5) and the terms containing the products of spin operator pairs $s^{\pm}$ and $s^z$, $s^+$ and $s^+$, $s^-$ and $s^-$ are neglected. Now, we move to the interaction representation by choosing the frame rotating with frequency $\Omega$ around the $z$ axis ($\rho_2 \to \rho_3 = u_3^+ \rho_2 u_3$, $u_3 = e^{-i\Omega t s^z}$). In this frame, we have $i\partial \rho_3 / \partial t = [H_3, \rho_3] + i\Lambda_2 \rho_3$,

where

$$H_3 = -\frac{1}{2}\omega_2 \sin\theta s^+ \left(e^{i\alpha t} e^{-i\psi} + e^{i(2\omega_{rf}+\alpha)t} e^{i\psi}\right) - h.c. + 2\omega_2 \cos\theta \cos(\omega_{rf} t + \psi) s^z, \qquad (6)$$

$\alpha = \Omega - \omega_{rf}$, in our case $|\alpha| \ll \Omega, \omega_{rf}$.

Rapidly oscillating ($e^{\pm i\omega_{rf} t}$, $e^{\pm i2\omega_{rf} t}$) terms in the Hamiltonian $H_3$ (6) can be eliminated by the Krylov–Bogoliubov–Mitropolsky method [31]. Assuming that $\Delta/\omega_{rf} \ll 1$ and averaging over the period $2\pi/\omega_{rf}$, we obtain the following effective Hamiltonian $H_{eff}$ up to the second order in $\omega_2/\omega_{rf}$: $H_3 \to H_{eff} = H_{eff}^{(1)} + H_{eff}^{(2)}$ with

$$H_{eff}^{(1)} = \langle H_3(t) \rangle = -(\omega_2/2)\sin\theta \left(s^+ e^{i(\alpha t - \psi)} + h.c.\right)$$

and $H_{eff}^{(2)} = (i/2)\langle [\int^t d\tau (H_3(\tau) - \langle H_3(\tau)\rangle), H_3(t)]\rangle = \Delta_{BS} s^z$.

In the above, the symbol $\langle ... \rangle$ denotes time averaging: $\langle A(t)\rangle = \frac{1}{T}\int_0^T A(t)dt$, where $T = 2\pi/\omega_{rf}$ and $\Delta_{BS} \approx \omega_2^2/4\omega_{rf}$ is the Bloch–Siegert-like frequency shift. After the canonical transformation $\rho_3 \to \rho_4 = u_4^+ \rho_3 u_4$, $u_4 = e^{i(\alpha t - \psi) s^z}$, the equation $i\partial \rho_3/\partial t = [H_{eff}, \rho_3] + i\Lambda_2 \rho_3$ is transformed into $i\partial \rho_4/\partial t = [H_4, \rho_4] + i\Lambda_2 \rho_4$ with $H_4 = (\alpha + \Delta_{BS})s^z - (1/2)\omega_2 \sin\theta(s^+ + s^-)$. This transformation eliminates the exponential dependence on time and the initial phase of the rf field in the interaction Hamiltonian $H_{eff}^{(1)}$.

The diagonalization of the Hamiltonian $H_4$ by means of the rotation operator $u_5 = e^{-i\xi s^y}$ ($\rho_5 \to \rho_5 = u_5^+ \rho_4 u_5$, $H_5 = u_5^+ H_4 u_5$, $\Lambda_5 = u_5^+ \Lambda_2 u_5$) yields

$$i\frac{\partial \rho_5}{\partial t} = [H_5, \rho_5] + i\Lambda_5 \rho_5, \qquad (7)$$

where $H_5 = \varepsilon s^z$, $\varepsilon = [\delta^2 + \omega_2^2 \sin^2\theta]^{1/2}$ is the frequency of the Rabi oscillations between the spin states dressed simultaneously by the MW and RF fields, $\delta = \Omega - \omega_{rf} + \Delta_{BS}$. Besides the diagonalization of the Hamiltonian $H_4$, $u_5$ transforms simultaneously the relaxation superoperator $\Lambda_2$ to the form of $\Lambda_5$.

We calculate $\Lambda_5 = u_5^+ \Lambda_2 u_5$ within the RWA approximation assuming that $\Gamma_\downarrow, \Gamma_\uparrow, \Gamma_\varphi \ll \omega_2$ and obtain

$$\Lambda_5 \rho_5 = \frac{\tilde{\Gamma}_\downarrow}{2} D[s^-]\rho_5 + \frac{\tilde{\Gamma}_\uparrow}{2} D[s^+]\rho_5 + \frac{\tilde{\Gamma}_\varphi}{2} D[s^z]\rho_5,$$

where

$$\tilde{\Gamma}_\downarrow = \frac{1}{4}(\Gamma_\downarrow + \Gamma_\uparrow)(1+\cos^2\xi) + \frac{1}{2}(\Gamma_\downarrow - \Gamma_\uparrow)\cos\xi + \frac{1}{4}\Gamma_\varphi \sin^2\xi,$$

$$\tilde{\Gamma}_\uparrow = \frac{1}{4}(\Gamma_\downarrow + \Gamma_\uparrow)(1+\cos^2\xi) - \frac{1}{2}(\Gamma_\downarrow - \Gamma_\uparrow)\cos\xi + \frac{1}{4}\Gamma_\varphi \sin^2\xi,$$

and

$$\tilde{\Gamma}_\varphi = \Gamma_\varphi \cos^2\xi + (\Gamma_\downarrow + \Gamma_\uparrow)\sin^2\xi, \quad \sin\xi = -\omega_2 \sin\theta/\varepsilon, \quad \cos\xi = \delta/\varepsilon.$$

Here $\tilde{\Gamma}_{\downarrow,\uparrow,\varphi}$ are the doubly-dressed relaxation rates of the doubly-dressed qubit states. These "dressed" states and relaxation rates arise due to taking into account the strong interaction between the qubit and the MW and RF fields.

The solution of equation (7) can be written as follows:

$$\rho_5(t) = e^{(-iL_5 + \Lambda_5)t}\rho_5(0). \tag{8}$$

The superoperator $L_5$ acts in accordance with the rule: $L_5 X = [H_5, X]$. The density matrix $\rho(t)$ in the laboratory frame is given by

$$\rho(t) = u_1 u_2 u_3 u_4 u_5 \rho_5(t) u_5^+ u_4^+ u_3^+ u_2^+ u_1^+, \tag{9}$$

where $\rho_5(t)$ is defined by equation (8) and $\rho_5(0) = u_5^+ u_4^+(0) u_2^+ \rho(0) u_2 u_4(0) u_5$; moreover, if the qubit is in the ground state, $\rho(0) = 1/2 - s^z$.

The following relations can be obtained:

$$e^{(-iL_5+\Lambda_5)t} s^\pm = e^{(\mp i\Omega - \tilde{\Gamma}_\perp)t} s^\pm, \quad e^{(-iL_5+\Lambda_5)t} s^z = e^{-\tilde{\Gamma}_\parallel t} s^z$$

and

$$e^{(-iL_5+\Lambda_5)t} a = a[1 + 2\mathrm{N}_{dd}(1 - e^{-\tilde{\Gamma}_\parallel t})s^z], \tag{10}$$

where $a = const$, $\mathrm{N}_{dd} = -(\tilde{\Gamma}_\downarrow - \tilde{\Gamma}_\uparrow)/\tilde{\Gamma}_\parallel = -(\gamma_{21} - \gamma_{12})\cos\theta\cos\xi/\tilde{\Gamma}_\parallel$,

$$\tilde{\Gamma}_\parallel = \tilde{\Gamma}_\downarrow + \tilde{\Gamma}_\uparrow = \Gamma_\parallel + (\Gamma_\perp - \Gamma_\parallel)\sin^2\xi, \quad \tilde{\Gamma}_\perp = (\tilde{\Gamma}_\downarrow + \tilde{\Gamma}_\uparrow + \tilde{\Gamma}_\varphi)/2 = \Gamma_\perp - (1/2)(\Gamma_\perp - \Gamma_\parallel)\sin^2\xi,$$

$$\Gamma_\parallel = \Gamma_\downarrow + \Gamma_\uparrow = \gamma_\parallel + (\gamma_\perp - \gamma_\parallel)\sin^2\theta, \qquad \Gamma_\perp = (\Gamma_\downarrow + \Gamma_\uparrow + \Gamma_\varphi)/2 = \gamma_\perp - (1/2)(\gamma_\perp - \gamma_\parallel)\sin^2\theta,$$

$$\gamma_\parallel = \gamma_{12} + \gamma_{21}, \quad \gamma_\perp = (\gamma_{12} + \gamma_{21} + \eta)/2.$$

The longitudinal, $T_1$, and transverse (coherence), $T_2$, relaxation times of the bare qubit in the laboratory frame are represented by the respective relaxation rates as $\gamma_\parallel = 1/T_1$ and $\gamma_\perp = 1/T_2$.

We observe that the doubly-dressed relaxation rates $\tilde{\Gamma}_\parallel$ and $\tilde{\Gamma}_\perp$ depend on a detuning of the MW frequency from the qubit resonant frequency as well as on a detuning $\delta$ between the radio frequency and $\omega_1$. At $\omega_2 = 0$ the expressions for $\tilde{\Gamma}_\parallel$ and $\tilde{\Gamma}_\perp$ reduce to those known for the singly dressed qubits [32] (see also [4,5]).

The density matrix $\rho_5$ describes the evolution of the qubit states doubly dressed by the MW and RF fields. This matrix is given by

$$\rho_5(t) = \frac{1}{2} + r(t)s^z + p(t)s^+ + p^*(t)s^-, \tag{11}$$

where

$$r(t) = \mathrm{N}_{dd}(1 - e^{-\tilde{\Gamma}_\parallel t}) + (\sin\theta\sin\xi\cos\psi - \cos\theta\cos\xi)e^{-\tilde{\Gamma}_\parallel t}$$

and

$$p(t) = \frac{1}{2}(\sin\theta\cos\xi\cos\psi + \cos\theta\sin\xi + i\sin\theta\sin\psi)e^{-i\varepsilon t}e^{-\tilde{\Gamma}_\perp t}. \tag{12}$$

Note that in the Hamiltonian $H_5 = \varepsilon s^z$ the value $\varepsilon$ is the energy (in frequency units) of each of four qubits arising due to dressing of the initial qubit by the MW and RF fields (Fig. 1). In this case, the operators $s^z$ and $s^\pm$ act in the space of the $|g\rangle$ and $|e\rangle$ vectors describing the ground and excited states of each of four qubits. From Eq. (11), we have $\rho_5^{st} = 1/2 + \mathrm{N}_{dd} s^z$ for the stationary state. Consequently, a stationary population difference of the doublet is $W_5^{st} = (\langle e|\rho_5|e\rangle - \langle g|\rho_5|g\rangle)/2 = \mathrm{N}_{dd}/2 = -(\tilde{\Gamma}_\downarrow - \tilde{\Gamma}_\uparrow)/2\tilde{\Gamma}_\parallel$. If the rate of the transition $\tilde{\Gamma}_\uparrow$ from the ground state of the upper doublet to the excited state of the lower doublet exceeds the rate $\tilde{\Gamma}_\downarrow$ of the inverse transition (see Fig. 1), the population inversion can be achieved. Since $\mathrm{N}_{dd} \sim \Delta\delta$, the inversion is realized at the opposite signs of detuning $\Delta$ and $\delta$.

The transitions between the levels within the doublet are forbidden as only the transitions between the levels of the different doublets can be realized. In the singly rotating frame (SRF), which rotates with frequency $\omega_{mw}$ around the $z$ axis of the laboratory frame, these transitions occur between the levels of the two doublets of the excited state $|2\rangle$ of the initial qubit as well as between the two doublets of its ground state $|1\rangle$ (Fig. 1).

The density matrix of the qubit in the SRF is given by

$$\rho(t) = u_1\left\{\frac{1}{2} + \frac{1}{4}\Big[e^{-i(\omega_{rf}t+\psi)}\Big(p(\cos\xi+1) + p^*(\cos\xi-1) + r\sin\xi\Big)\Big(s^+(\cos\theta+1) + \right. \tag{13}$$

$$+ s^-(\cos\theta - 1) - 2s^z \sin\theta) + h.c.] + (r\cos\xi - (p + p^*)\sin\xi)\left(s^z \cos\theta + \frac{1}{2}(s^+ + s^-)\sin\theta\right)\right\} u_1^+,$$

where $p$ and $r$ is defined by Eq. (12).

By using the density matrix (13) one can calculate observables of the qubit for arbitrary time. In particular, the population difference, the absorption and dispersion signals and the correlation functions (upon additionally using the regression theorem [33]) can be obtained.

### III. Rabi oscillations on doubly-dressed spin states

In conventional time-resolved EPR or NMR experiments the absorption signal proportional to the projection of the magnetization of the qubit onto the $y$ axis is usually observed [11,13]. Using equation (13), in the SRF, the absorption signal can by written as:

$$V_{SRF} = \frac{1}{2i}\left(\langle 1|\rho_{SRF}|2\rangle - \langle 2|\rho_{SRF}|1\rangle\right) = \frac{1}{2} N_{dd} \sin\xi(1 - e^{-\tilde{\Gamma}_\| t})\sin(\omega_{rf} t + \psi) +$$

$$+ \frac{1}{4}\left\{\sin\theta\sin^2\xi\left(\sin\omega_{rf} t + \sin(\omega_{rf} t + 2\psi)\right) - 2\cos\theta\sin\xi\cos\xi\sin(\omega_{rf} t + \psi)\right\} e^{-\tilde{\Gamma}_\| t} +$$

$$+ \frac{1}{8}\left\{\sin\theta\left[(\cos\xi + 1)^2 \sin(\omega_{rf} + \varepsilon)t + (\cos\xi - 1)^2 \sin(\omega_{rf} - \varepsilon)t - \right.\right.$$

$$-\sin^2\xi\left(\sin\left((\omega_{rf} + \varepsilon)t + 2\psi\right) + \sin\left((\omega_{rf} - \varepsilon)t + 2\psi\right)\right)\right] +$$

$$+ 2\cos\theta\sin\xi\left((\cos\xi + 1)\sin\left((\omega_{rf} + \varepsilon)t + \psi\right) + (\cos\xi - 1)\sin\left((\omega_{rf} - \varepsilon)t + \psi\right)\right)\right\} e^{-\tilde{\Gamma}_\perp t}. \tag{14}$$

In such a case the density matrix is $\rho_{SRF} = u_2 u_3 u_4 u_5 \rho_5(t) u_5^+ u_4^+ u_3^+ u_2^+$). The relaxation rates can be expressed as

$$\tilde{\Gamma}_\| = \gamma_\| + (\gamma_\perp - \gamma_\|) f(\theta, \xi), \quad \tilde{\Gamma}_\perp = \gamma_\perp - \frac{1}{2}(\gamma_\perp - \gamma_\|) f(\theta, \xi), \tag{15}$$

where $f(\theta, \xi) = \sin^2\theta + (1 - \frac{3}{2}\sin^2\theta)\sin^2\xi$.

It is follows from Eq. (14) that in the SRF the Rabi oscillations at three frequencies, $\omega_{rf}$ and $\omega_{rf} \pm \varepsilon$, are realized. These oscillations result from the allowed quantum multiphoton transitions between states of four qubits formed by the double dressing of an initial qubit with the bichromatic field [34].

The predicted frequencies of the Rabi oscillations between the doubly dressed states were confirmed experimentally [10,12,19] including the Bloch-Siegert effect [11,18,35]. The amplitude of the Rabi oscillations at each frequency is strongly dependent on the RF field phase. Such a dependence was analyzed in ref. [34] and was observed in ref. [36] for $\delta = 0$. Dephasing and relaxation of the doubly dressed states have not been studied experimentally in details untill now.

Figures 2 and 3 illustrate the behavior of the Rabi oscillations between the doubly dressed states versus the detuning $\delta$. We average a large number of the oscillations assuming that the RF phase is random. Figure 4 shows the Fourier spectra of the Rabi oscillations for $\delta = 0$ as a function of the fixed RF phase. The relaxation parameters of defects such as the isolated neutral substitutional nitrogen (P1 center) and the negatively charged nitrogen vacancy (NV) center in diamonds were used in our calculations. The latter centers are particularly promising quantum systems with wide applications in physics and biology [37]. Their spin coherence time $T_2$ depends on the concentration of paramagnetic centers in the diamond lattice and can be changed from a few microseconds in nitrogen-rich type-Ib diamond to a few milliseconds in ultrapure diamonds [37-39]. In the latter case, the coherence time $T_2$ becomes comparable with the spin-lattice relaxation time $T_1$ [19,40]. Due to a long coherence time at room temperature, they are among the most promising candidates for solid-state qubits [37,41-43]. On the other hand, these centers can be used to study coherent properties of qubits in a wide range of relaxation parameters, from $T_2 \ll T_1$ to $T_2 = 2T_1$.

According to Eqs. (14) and (15), not only the frequencies and amplitudes of the Rabi oscillations, but also the oscillation decay rates strongly depend on detuning $\delta$. Figure 5 displays the decay rates $\tilde{\Gamma}_\parallel$ and $\tilde{\Gamma}_\perp$ for the oscillations at the central, $\omega_{rf}$, and side, $\omega_{rf} \pm \varepsilon$, frequencies, respectively. For $\delta = 0$, the decay rate $\tilde{\Gamma}_\parallel$ has its minimum value and the decay rate $\tilde{\Gamma}_\perp$ has its maximum value. With increasing detuning, the decay rate $\tilde{\Gamma}_\parallel$ of the oscillations at the central frequency increases but the decay rate $\tilde{\Gamma}_\perp$ of the oscillations at the side frequencies decreases. There is the value for the positive and negative detuning at which the decay rates at the central and side frequencies become equal. For other magnitudes of detuning, the decay of the observed signals is described by the two-exponential function.

The situation presented in Figs. 2 – 4 and in Fig. 5a is typical for time-resolved EPR and NMR experiments for an ensemble of spin qubits in solids when $T_2 = 1/\gamma_\perp \ll T_1 = 1/\gamma_\parallel$. Note that at $T_2 \ll T_1$ the decay rate of the Rabi oscillations in the spin states dressed only by the MW field increases with an increase in the detuning $\Delta = \omega_0 - \omega_{mw}$ [4,32]. At the same time, for the doubly dressed states at $\Delta = 0$, an increase in the detuning $\delta$ prolongs the decay of Rabi oscillations at frequencies $\omega_{rf} \pm \varepsilon$ and simultaneously increases their decay at frequency $\omega_{rf}$. For an ensemble of spin qubits it is difficult to fulfill the condition $T_2 > T_1$ and to achieve a crossover to the opposite dependence of the decay of Rabi oscillations on the detuning $\delta$ (Fig. 5b). However, this condition can be realized for single qubits, for instance for nitrogen-vacancy spin qubits [30] or two-level semiconductor InAs/GaAs quantum dots [7].

**IV. Steady-state population inversion of the doubly dressed levels**

In the stationary regime, the absorption signal (14) in the SRF can be written in the form:

$$V_{SRF}^{st} = \frac{1}{2} N_{dd} \sin \xi \sin(\omega_{rf} t + \psi) \equiv V \sin(\omega_{rf} t + \psi), \qquad (16)$$

where

$$V = \frac{\gamma_{21} - \gamma_{12}}{2\tilde{\Gamma}_{\parallel}} \frac{\Delta \delta \omega_1 \omega_2}{\varepsilon^2 \Omega^2}. \qquad (17)$$

Note that, in particular, the continuous wave (cw) bichromatic field under consideration is commonly used in cw EPR spectroscopy [44]. RF modulation of the static magnetic field with subsequent phase-sensitive detection at the radio frequency improves the signal-to-noise ratio. Eq. (17) describes the absorption EPR signal observed in such experiment at the $90^0$-out-of phase detection. Fig. 6 shows $V$ as a function of detuning of MW and radio frequencies $\Delta$ and $\delta$ at the fixed value of the MW amplitude. We see that when the sings of detuning $\Delta$ and $\delta$ are different, $V$ becomes negative and the inversion of the EPR absorption signal is achieved. This inversion is a signature of a steady-state population inversion of the doubly dressed-state levels. Fig. 7 demonstrates the influence of the relaxation rates on the magnitude of the steady-state population inversion. We observe that the maximal inversion is achieved at $T_2 = 2T_1$.

In accordance with Eq. (17), the inversion of the absorption signal $V$ observed at the $90^0$-out-of-phase detection can be used to reveal the predicted steady-state population inversion $N_{dd} = -(\tilde{\Gamma}_{\downarrow} - \tilde{\Gamma}_{\uparrow})/\tilde{\Gamma}_{\parallel}$. Because cw EPR experiments are usually performed at a fixed radio frequency, the population inversion can be manifested as an inversion of the EPR signal when changing the value of $\omega_1$ results in the change of the sign of detuning $\delta$ near $\omega_{rf} = \omega_1$. This effect can be observed in the dependence of the EPR absorption on the MW power for samples with long electron spin relaxation times. Fig. 8 illustrates the inversion of $V$ for the parameters used in the experiment for the P1 centers in high-quality diamonds [45]. The relaxation times measured by cw saturation methods in the experimental samples were $T_2 \geq 0.87$ μs, $T_1 \leq 2.7$ ms. The estimated concentration of the P1 centers in the investigated sample was about $4 \times 10^{20}$ m$^{-3}$. The time-resolved EPR measurements have shown that spin-spin relaxation time $T_2$ at such concentration of the P1 centers is longer than one millisecond achieving the value of $T_1$ [38,40]. Therefore, we assumed that $T_2 = T_1$ in the calculations presented in Fig. 8. Because the range of $\Delta$ used in our calculations is limited by the condition $\Delta \ll \omega_1$, only the central part of the observed EPR signal is described. Nevertheless, it is enough to reveal the main feature in the dependence of the EPR absorption on the MW power, namely the inversion of this signal when an increase in $\omega_1$ changes the sign of detuning $\delta$. Therefore, we conclude that the inversion of the out-of-phase EPR signal observed in ref. [45] reveals the predicted steady-state population inversion of the doubly dressed levels. The observed inversion of the EPR signal has not been explained until now.

### V. Conclusions

We have shown that the relaxation of the doubly dressed states depends on detuning of the radio frequency from the MW-field Rabi frequency. The character of the dependence is insensitive to the phase RF field and is determined by the ratio of $T_1$ and $T_2$. Detuning prolongs the decay of the Rabi oscillations between some doubly dressed states and simultaneously increases their decay between the other states. We have shown that a steady-state population inversion of the doubly dressed levels can be achieved. Our predictions have been demonstrated for NV and P1 centers in diamond.

Besides EPR and NMR, our results are potentially applicable to optimization of single-spin magnetometers, emulations of a hybrid spin-mechanical system as well as amplification due to doubly dressed-state inversion.

**Figure legends**

Fig. 1. The energy level diagram showing the singly and doubly dressed levels, transition frequencies (solid arrows) and relaxation rates (dashed arrows). The transitions between the double dressed levels are presented in the frame, which rotates with frequency $\omega_{mw}$ around the $z$ axis of the laboratory frame.

Fig. 2. The Rabi oscillations on dressed spin states at $\omega_1 = 2\pi$ 1.00 MHz, $\Delta = 0$, $T_2 = 10$ μs, $T_1 = 2$ ms. (a): $\omega_2 = 0$. (b): $\omega_2 = 2\pi$ 0.2 MHz, $\omega_{rf} = 2\pi$ 1.0 MHz. (c): $\omega_2 = 2\pi$ 0.2 MHz, $\omega_{rf} = 2\pi$ 0.85 MHz. (d), (e) and (f): the Fourier spectra of the signals shown in (a) - (c). The inset shows the sidebands for $\delta = 0$ and $\delta = -2\pi$ 0.15 MHz shifted and rescaled to lie on top of each other.

Fig. 3. The Fourier spectra of the Rabi oscillations as a function of detuning $\delta$. Other parameters are the same as in Fig.2. The dashed line shows the position of the spectrum presented in Fig. 2e. The dotted line shows the position of the spectrum presented in Fig.2f.

Fig. 4. The Fourier spectra of the Rabi oscillations at $\delta = 0$ versus the RF phase. Other parameters are the same as in Fig.2.

Fig. 5. The decay rates for the Rabi oscillations at the central (solid line) and side (dashed line) frequencies as a function of detuning $\delta$. $\Delta = 0$, $\omega_1 = 2\pi$ 1.00 MHz, $\omega_2 = 2\pi$ 0.2 MHz. (a) $T_2 = 10$ μs, $T_1 = 2$ ms (b) $T_2 = 4$ ms, $T_1 = 2$ ms.

Fig. 6. Amplitude of the cw EPR signal $V$ versus detuning of MW and radio frequencies $\Delta$ and $\delta$. $\omega_2 = 2\pi$ 0.2 MHz, $\omega_1 = 2\pi$ 1.00 MHz. $T_1/T_2 = 0.5$.

Fig. 7. Amplitude $V$ as a function of detuning $\Delta$ for the fixed value of $\delta = -2\pi$ 0.15 MHz shown the dashed line in Fig. 6, $T_1/T_2 = 0.5$ (solid line); $T_1/T_2 = 1$ (dashed line) $T_1/T_2 = 10$ (dotted line). Other parameters are the same as in Fig. 6.

Fig. 8. Inversion of the out-of-phase EPR signal at an increase in the MW power. $\omega_{rf} = 2\pi$ 0.1 MHz; $\omega_1/2\pi = 0.12$ MHz (solid line) and 0.04 MHz (dashed line), $\omega_2 = 2\pi$ 0.028 MHz, $T_2 = T_1 = 2$ ms.

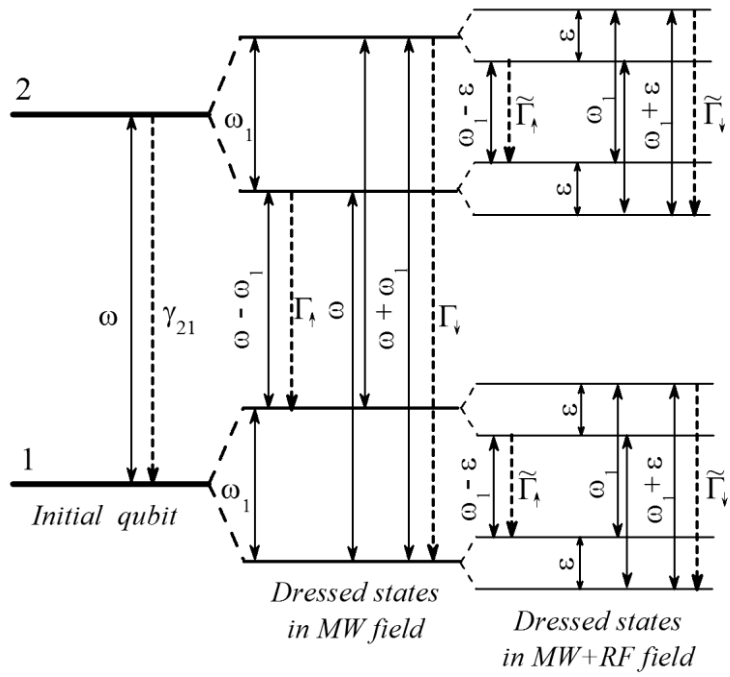

Fig. 1.

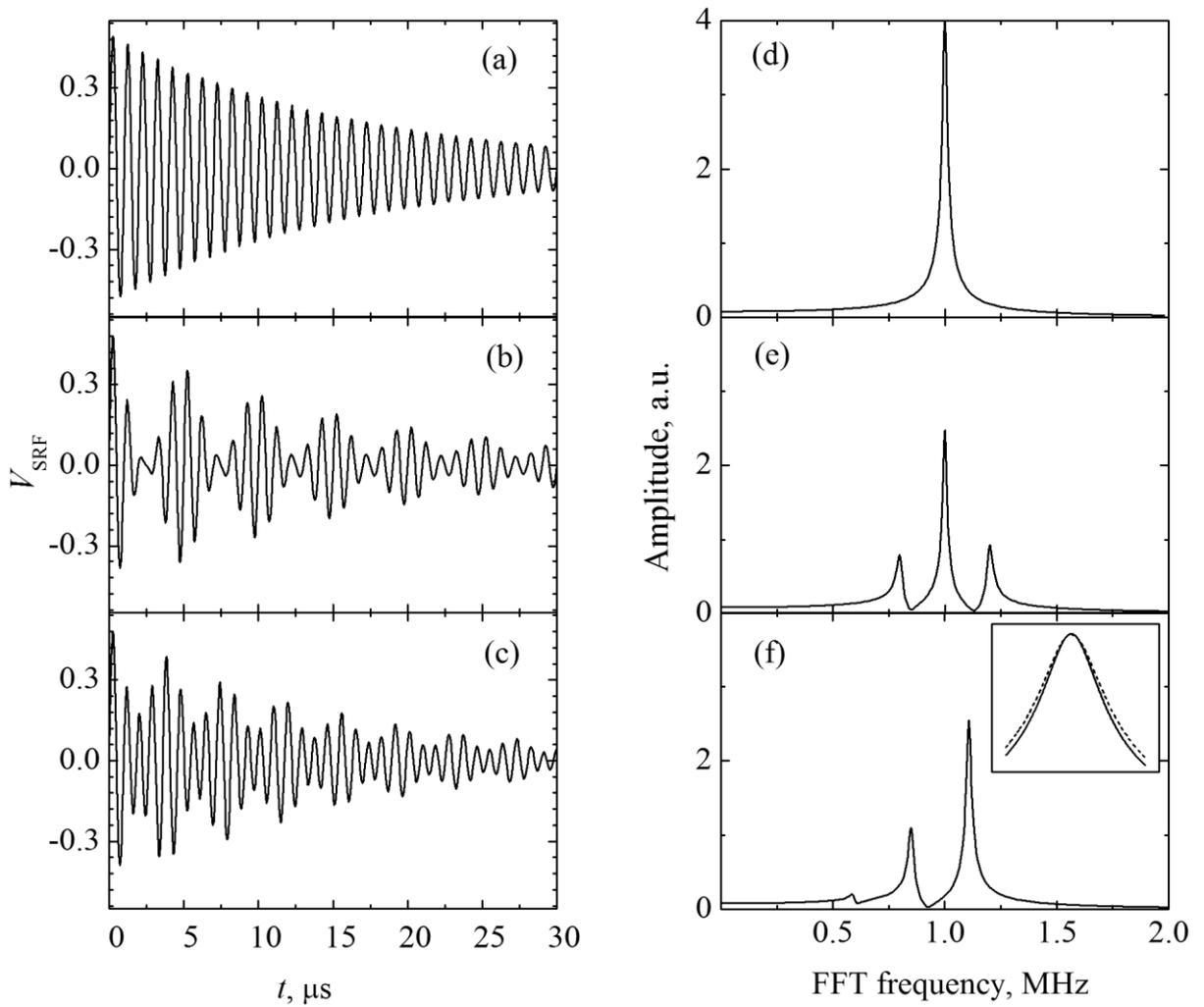

Fig. 2.

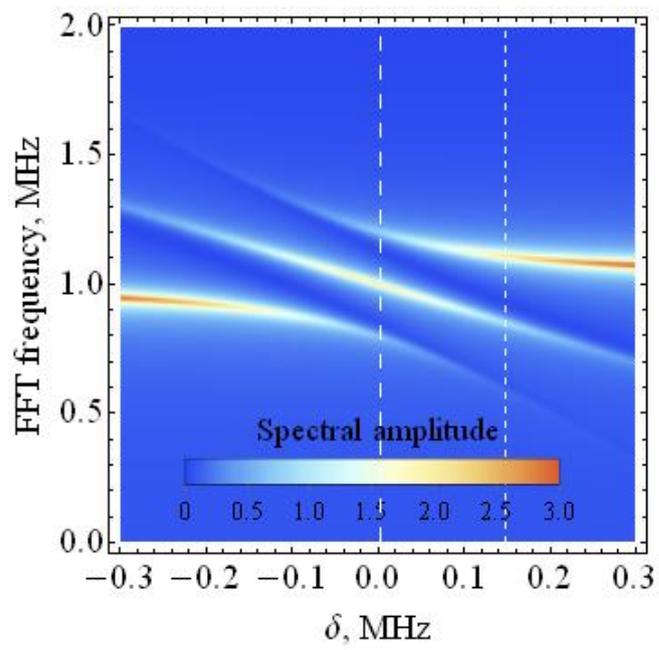

Fig. 3.

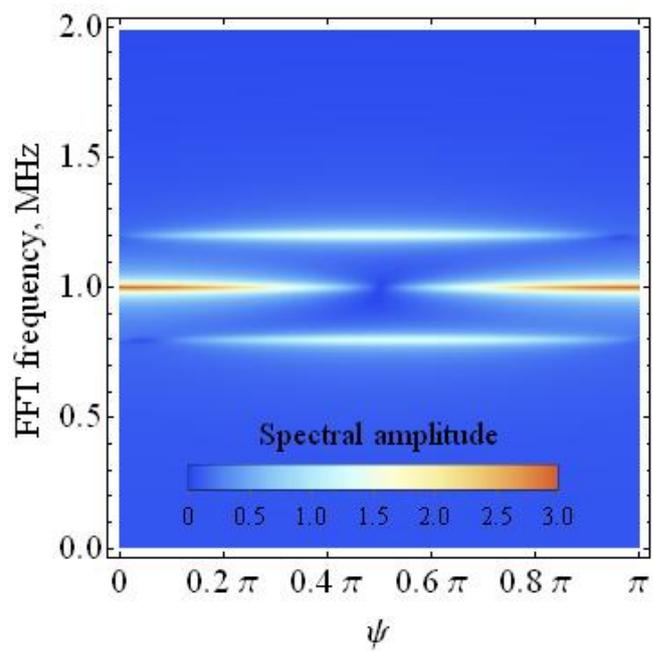

Fig. 4.

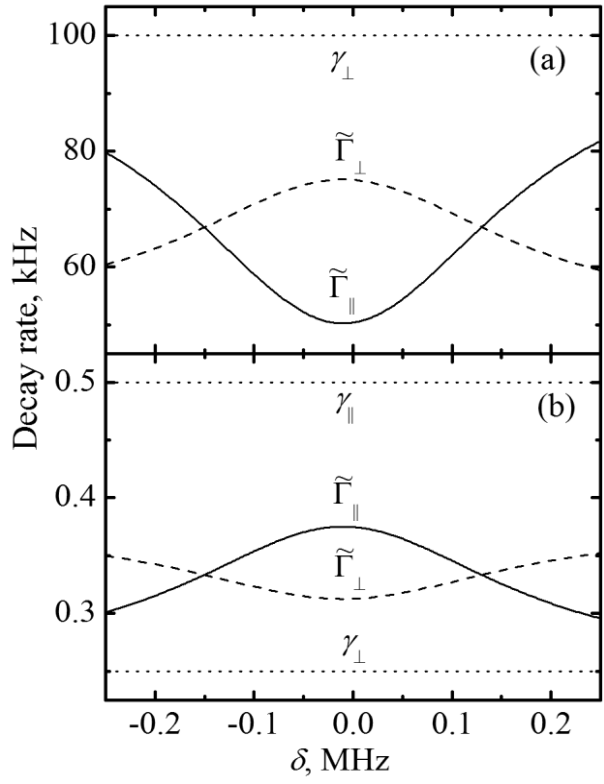

Fig. 5.

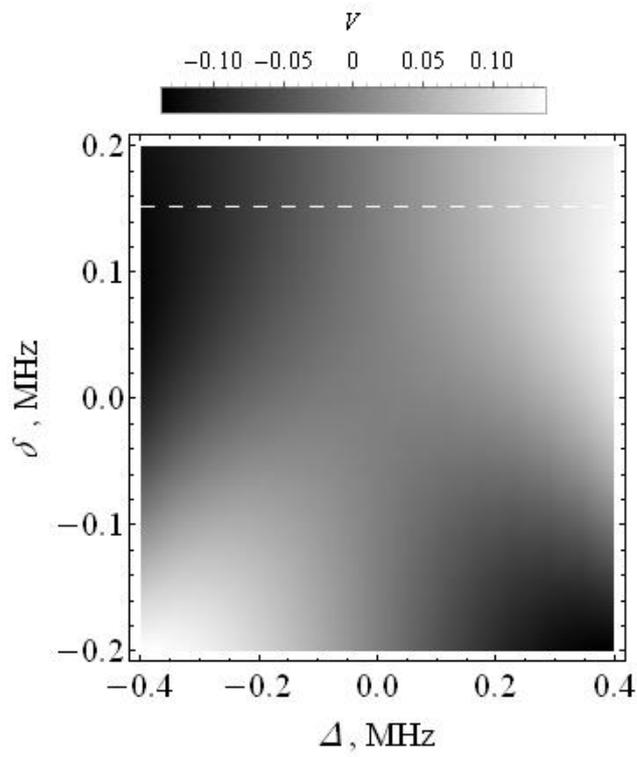

Fig. 6.

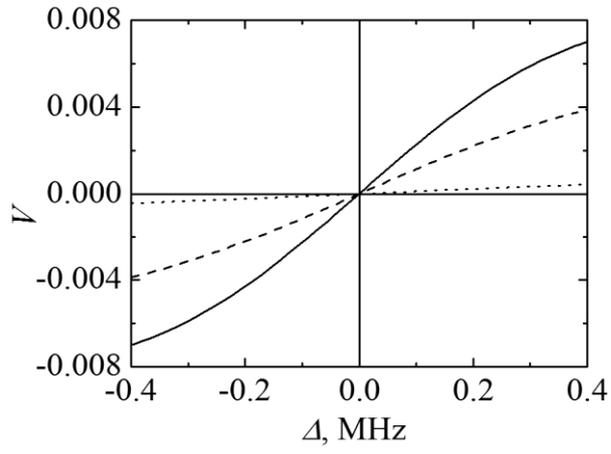

Fig. 7.

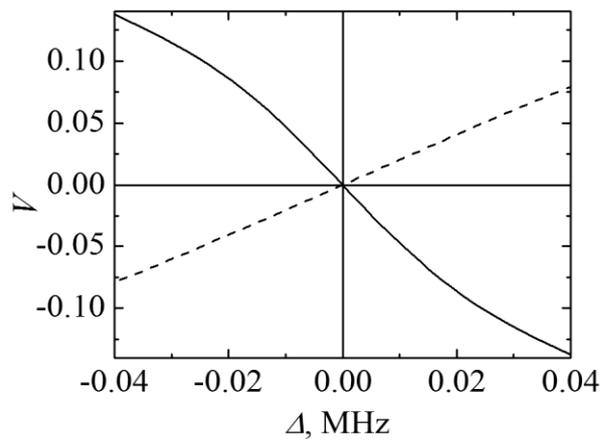

Fig. 8.